\documentclass[9pt,twocolumn,twoside]{opticajnl}
\journal{opticajournal} 

\setboolean{shortarticle}{true}

\usepackage{comment}
\usepackage{graphicx}

\title{Experimental and on-sky demonstration of spectrally dispersed wavefront sensing using a photonic lantern}

\author[a,*]{Jonathan Lin}
\author[a]{Michael P. Fitzgerald}
\author[b]{Yinzi Xin}
\author[a]{Yoo Jung Kim}
\author[c,d]{Olivier Guyon}
\author[e]{Barnaby Norris}
\author[f]{Christopher Betters}
\author[f]{Sergio Leon-Saval}
\author[c,g]{Kyohoon Ahn}
\author[c]{Vincent Deo}
\author[c]{Julien Lozi}
\author[c,h]{Sébastien Vievard}
\author[i]{Daniel Levinstein}
\author[i]{Steph Sallum}
\author[b]{Nemanja Jovanovic}

\affil[a]{University of California Los Angeles, Physics \& Astronomy Department, 475 Portola Plaza, Los Angeles, USA, 90095}

\affil[b]{California Institute of Technology, Department of Astronomy, Pasadena, CA 91125, USA}

\affil[c]{National Astronomical Observatory of Japan, Subaru Telescope, 650 North Aohoku Place, Hilo, HI 96720, USA}

\affil[d]{The University of Arizona, Department of Astronomy and Steward Observatory, 933 N. Cherry Ave., Tucson, AZ 85719, USA}

\affil[e]{The University of Sydney, Sydney Institute for Astronomy, Physics Road, Sydney, NSW 2006, Australia}

\affil[f]{The University of Sydney, Sydney Astrophotonic Instrumentation Laboratory, Sydney, NSW 2006, Australia}

\affil[g]{Korea Astronomy and Space Science Institute, 776 Daedeok-daero, Yuseong District, Daejeon, South Korea}

\affil[h]{Astrobiology Center, 2-21-1, Osawa, Mitaka, Tokyo, 181-8588, Japan}

\affil[i]{University of California, Irvine, Department of Physics \& Astronomy, 4129 Frederick Reines Hall, Irvine, CA 92697, USA}
\affil[*]{jon880@astro.ucla.edu}

\begin{abstract}
Adaptive optics systems are critical in any application where highly resolved imaging or beam control must be performed through a dynamic medium. Such applications include astronomy and free-space optical communications, where light propagates through the atmosphere, as well as medical microscopy and vision science, where light propagates through biological tissue. Recent works have demonstrated common-path wavefront sensors for adaptive optics using the photonic lantern, a slowly varying waveguide that can efficiently couple multi-moded light into single-mode fibers. We use the SCExAO astrophotonics platform at the 8-m Subaru Telescope to show that spectral dispersion of lantern outputs can improve correction fidelity, culminating with an on-sky demonstration of real-time wavefront control. To our best knowledge, this is the first such result for either a spectrally dispersed or a photonic lantern wavefront sensor. Combined with the benefits offered by lanterns in precision spectroscopy, our results suggest the future possibility of a unified wavefront sensing spectrograph using compact photonic devices.
\end{abstract}

\setboolean{displaycopyright}{false} 

\begin{document}

\maketitle

\section{Introduction}
\label{sect:intro}  

Adaptive optics (AO) systems cancel wavefront distortions in real-time by driving wavefront correctors such as deformable mirrors (DMs) with wavefront sensors (WFSs), typically in closed-loop control. Such systems have found use in a wide range of imaging and telecommunications applications, including ground-to-space optical communications \cite{Tyson}, deep tissue microscopy \cite{rodriguez}, and remote sensing \cite{Daigle}. 
In astronomy, AO systems correct for wavefront distortions which may originate from the overhead turbulent atmosphere (in the case of ground-based observation) or time-varying imperfections in the scientific instrument. AO has enabled highly-resolved imaging of the galactic center \cite{sag}, exoplanets \cite{hr8799}, and more; \cite{oli} provides a review. 

One of the primary goals for astronomy for the next decades is the direct imaging of an Earth-like exoplanet and the identification of potential biosignatures \cite{decadal}. This achievement will require the separation of exoplanetary light from that of a host star which outshines its companion by 10 orders of magnitude in visible wavelengths \cite{Traub:10}, for instance by using coronagraphy \cite{cor}. However, coronagraphic contrast is highly sensitive to wavefront distortions, requiring correction at a precision and stability which exceeds the capabilities of current AO systems. Such systems typically use general-purpose WFSs such as the Shack-Hartmann lenslet array or pyramid WFS, which re-image the telescope pupil and split off a portion of the light to a dedicated WFS arm. This gives rise to non-common-path aberrations (NCPAs): quasi-static aberrations arising from opto-mechanical deformations which affect one arm of the instrument but not the other \cite{Martinez:12:NCPA1}. Other sources of wavefront error include the low-wind effect \cite{sauvage} and petaling \cite{petal}, both of which arise from fragmentation of the telescope aperture and induce discontinuous phase aberrations over the pupil.
\begin{figure*}
  \centering
  \includegraphics[width=0.8\textwidth]{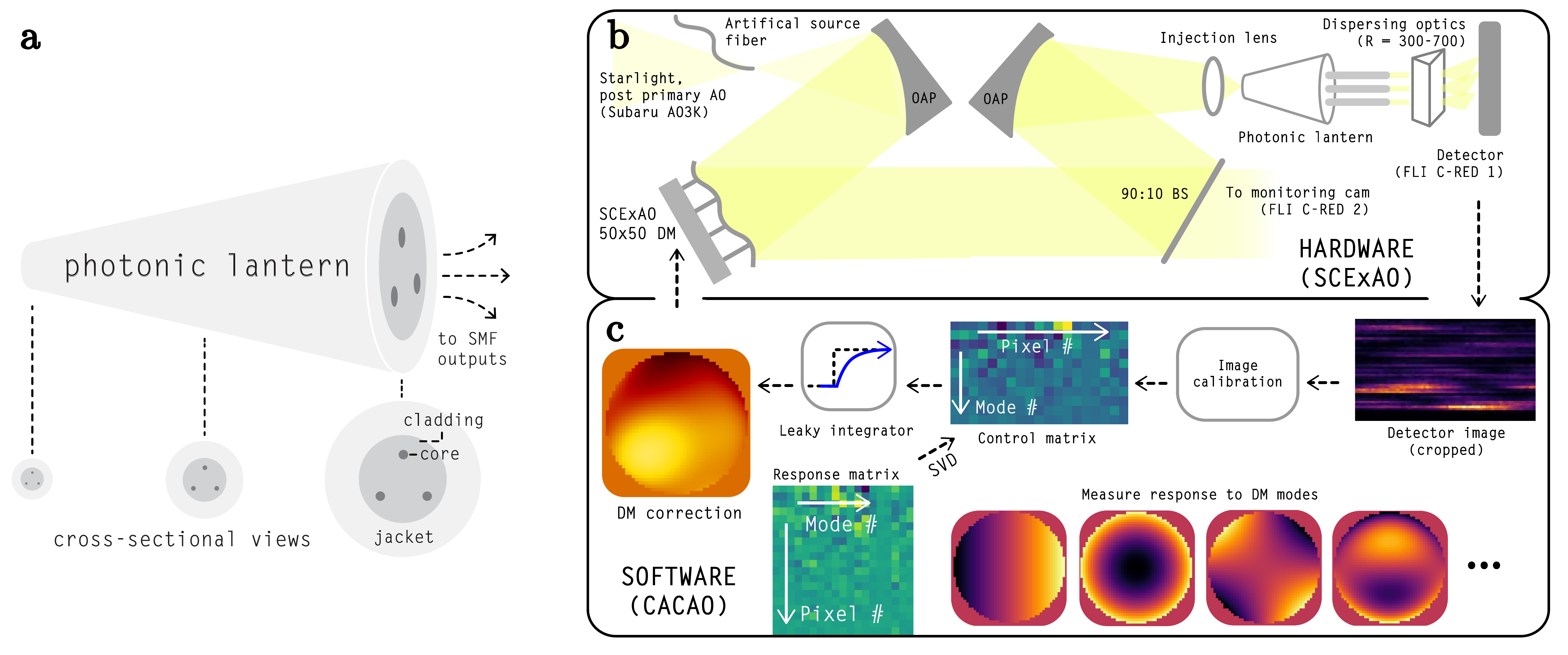}
  \caption{\textbf{a:} A 3-port PL, with 3 outputs. The PL couples light copropagating in multiple fiber modes into multiple single-mode outputs. Note that we use a 19-port PL; a 3-port PL is drawn for simplicity. \textbf{b:} Simplified beam diagram of the astrophotonics platform at SCExAO. Light comes from either the facility AO or a supercontinuum laser, and reflects off a $50\times 50$ actuator DM. The light is then divided between an internal camera, used to monitor the Strehl ratio, and the PL, which is dispersed at a resolution of several hundred. \textbf{c:} Software steps used to perform wavefront control; we use a linear phase retrieval method. }\label{fig:pl}
\end{figure*}
While such aberrations are difficult to correct with pupil-reimaging WFSs, they can be sensed from the focal plane: a strategy that also mitigates NCPAs because the instrument now has a single common path. To date, several focal-plane sensing techniques have been proposed (e.g. DRWHO \cite{DRWHO}), as well as dedicated sensor designs like the self-coherent camera \cite{scc} and mode-sorting approaches using holographic phase masks \cite{andersen,zepp}. Recently, focal-plane sensing was also demonstrated with a photonic lantern (PL; \cite{Leon-Saval:05}). As shown in Figure \ref{fig:pl}a, the PL is a slowly varying waveguide similar to a tapered multicore fiber which can efficiently couple aberrated telescope light into single-mode fibers (SMFs); optical throughputs of PLs have reached 97\% in laboratory testing \cite{Noord}. Such devices also have potential astronomical uses beyond AO: for instance, they can stably inject light into spectrographs \cite{Lin:21,vispl} e.g. for high-resolution exoplanet spectroscopy, probe sub-diffraction-limit spatial features with spectroastrometry~\cite{Kim:24}, and null starlight for high-contrast imaging \cite{Xin}. Simultaneously, the PL can drive a second stage of common-path wavefront correction \cite{lin:22}, complementing the telescope's primary AO.

Previous demonstrations of the PL WFS used the non-dispersed spot images of a PL's single-mode outputs to perform phase retrieval \cite{barnaby} and drive wavefront correction~\cite{Lin:23}. The purpose of this work is to show that the sensing capability of the PL is retained, and even improved, when wavelength dispersion is introduced, and that such devices are suitable for second stage correction of NCPA and low-wind effect. To this end, we use the SCExAO high-contrast imaging testbed at the 8 m Subaru Telescope \cite{scex}, also used in \cite{Lin:23}. This testbed contains an astrophotonics platform that was recently upgraded to include a low-resolution diffraction-limited spectrograph, as well as a 19-port PL optimized for high throughput in near-infrared wavelengths ($\lambda = 1-1.8 \, \mu$m). This device was fabricated at the Sydney Astrophotonics Instrumentation Laboratory by tapering a multicore fiber with 19 cores. We present both off-sky and on-sky demonstrations of real-time wavefront control.

\section{Results}\label{results}
\subsection{Off-sky laboratory demonstration}
We first present off-sky results from SCExAO. Figure \ref{fig:pl}b gives an overview of our setup: briefly, this testbed emulates a point source using a supercontinuum laser, which is collimated onto a 2500 actuator DM. 90\% of the light is directed to a 4-axis stage which holds the PL in a focal plane and enables tuning of alignment and focal ratio, while the remainder is sent to an FLI C-RED 2 detector for PSF and Strehl ratio monitoring. The output of the PL is dispersed at a resolution of 700 - 300 over $\lambda=1-1.8$ $\mu$m with a prism onto an FLI C-RED 1 detector. When on-sky, the laser source is replaced with starlight from the facility AO system (AO3K, \cite{julienao3k}), which uses a separate 3228-actuator DM and pyramid WFS. See Supplement 1 for more details.

Figure \ref{fig:pl}c overviews the calibration and control process. First, we measure the slope of the PL's response to the first 100 non-piston Zernike modes. This gives the response matrix, which approximates the relation between the pupil phase and WFS output; a singular value decomposition (SVD) of this matrix determines our control modes. To compute the phase correction, the phase retrieved by the PL WFS is fed into a leaky integrator. We then closed the AO loop (i.e. applied the correction to the DM in real-time) at a framerate of 1.5 kHz. After running a ``self-test'' of the loop, where a static amount of each control mode is injected into the system one-by-one on a separate channel of the DM, we found that the first 52 control modes --- each a linear combination of Zernike modes --- were clearly corrected, with correction of the remainder too slow to be useful. Note that an un-dispersed 19-port PL senses strictly less than 19 aberration modes \cite{lin:22}, and that previous experiments from \cite{Lin:23} demonstrated control over only 15 modes. We provide an explanation for this increase in \S\ref{disc}. Next, to test the loop, we generated phase screens with an RMS amplitude of 150 nm and windspeed of 10 m/s, and applied them to an independent channel of the DM. We used high-pass-filtered Kolmogorov phase screens to emulate first-stage correction of atmospheric turbulence by Subaru's facility AO. Finally, we recorded the amplitudes for each control mode, as measured through the PL WFS, in both open- and closed-loop operation. The ratio of the closed-loop and open-loop power spectral densities (PSDs) estimates the squared modulus of the rejection transfer function. We find that the correction loop is stable up to a limit of 150 - 200 nm RMS of wavefront error, dependent on alignment and control parameters. This somewhat limited dynamic range is expected by simulations, e.g. \cite{lin:22}.

\begin{figure}
    \centering
    \includegraphics[width=\columnwidth]{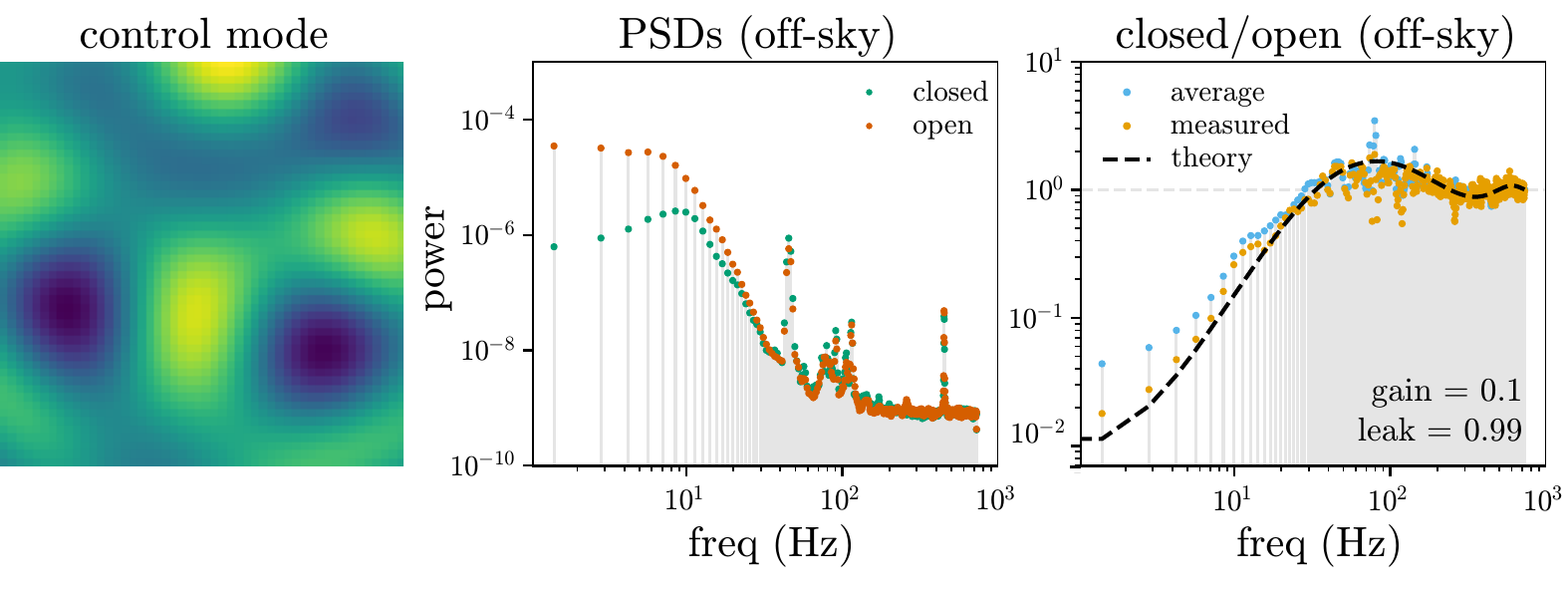}
    \caption{Wavefront control performance of the dispersed PL WFS during daytime laboratory tests. Left: one of the control modes. Middle: Open- and closed-loop PSDs of this mode's amplitude. Right: the ratio of the closed and open loop PSDs, which estimates the squared modulus of the rejection transfer function. Orange dots correspond to the given mode, and blue to the average over all modes; the dashed black lines give the expected transfer function for our control parameters. }
    \label{fig:3}
\end{figure}

Figure \ref{fig:3} shows the PSDs and rejection transfer function for an example control mode of the Zernike-calibrated 19-port PL WFS. We find good agreement with an analytic model derived in \citenum{Lin:23}, except at the slowest frequencies where the deviation approaches 2-3$\times$. This suggests that a subset of control modes had lower sensitivity and hence slower correction than expected; one possible cause is a slow optomechanical drift within SCExAO, which would change the alignment of the PL and eventually outdate the sensor calibration. See Supplement 1 for a similar test where the PL was calibrated in zonal (DM actuator) basis. 
\subsection{On-sky demonstration}
On 9/17/2024, we demonstrated closed-loop wavefront control using the spectrally dispersed 19-port PL at Subaru/SCExAO, using Humu (Altair) as our natural guide star. This demonstration was performed as part of SCExAO engineering time, under proposal ID S24B-EN03 (PI Julien Lozi). The seeing during the test was $\sim 0.5$", as estimated by the Maunakea Weather Center. Our test was also downstream of AO3K, which provided first-stage wavefront correction and yielded $\sim$50\% Strehl for $\lambda\sim 1.5-1.8$ $\mu$m, as measured from the PSF monitoring camera: as such, this demonstration is also the first to show the PL working as a second-stage focal plane WFS.

\begin{figure*}
    \centering
    \includegraphics[width=0.8\textwidth]{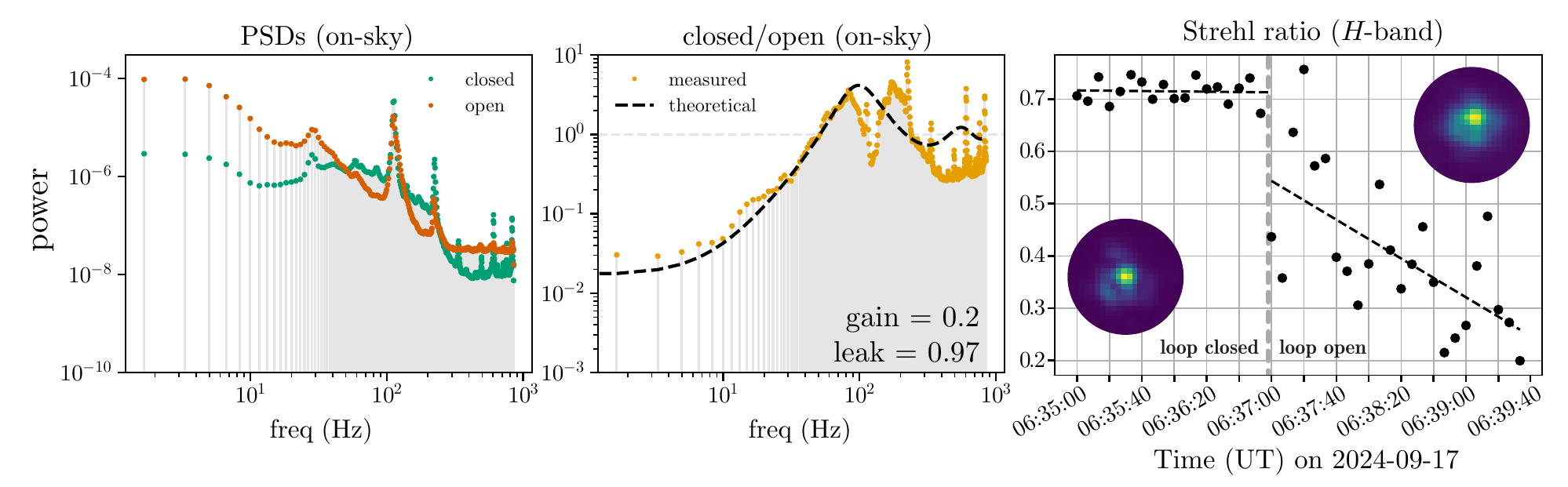}
    \caption{Left: on-sky open- and closed-loop PSDs for the dispersed 19-port PL WFS. Middle: the ratio of the PSDs, which estimates the squared modulus of the rejection transfer function. Agreement between the experimentally measured transfer function and the theoretical expectation for our control parameters is good up to around 100 Hz.  Right: the $H$-band Strehl ratio as a function of time. The wavefront control loop is initially closed and is then opened. Dashed black lines show linear regressions for the open- and closed-loop Strehl data. Circular insets show the averaged $H$-band PSF in under open- and closed-loop operation.}\label{fig:onsky}
\end{figure*}

Because our PL's spectral traces overlap on the detector, we were unable to account for spectral differences between SCExAO's calibration source and that of our natural guide star, and had to calibrate the WFS on-sky. This overlap is due to hardware constraints, which will be rectified in the future with a new PL; for more details, see Supplement 1. In the interest of on-sky time, we only calibrated the WFS against the first 30 non-piston Zernikes. The SVD of this matrix obtained 14 control modes, though higher fidelity control should be possible using an on-sky calibration with more basis modes or an off-sky, daytime calibration, as in the previous section. Finally, we closed the loop using a leak of 0.97 and a gain of 0.2 at a rate of 1.7 kHz and latency of 3.77 frames, using the PL and SCExAO DM to perform secondary correction after AO3k. Our somewhat high latency could be reduced with improvements to the (currently Python-based) processing of the dispersed PL image. Figure \ref{fig:onsky} gives the average open- and closed-loop PSDs for the 14 control modes, as well as the estimated rejection transfer function; the peak at 100 Hz is a result of noise amplification at high gain, and is captured by the analytic model. Beyond 100 Hz, we find deviations which might be related to the large spikes in the PSDs near 100 and 200 Hz, perhaps corresponding to strong vibrations. 

The rightmost panel of \ref{fig:onsky} plots the estimated $H$-band Strehl ratio over the course of the test, and includes insets of the $H$-band point spread function (PSF) immediately before and after loop closing, averaged over a two minute timescale. Note that the closed-loop data was taken before the open-loop data, and that after opening the loop, the seeing gradually increased to $\sim$1", at which point we were not able to close the PL loop again due to the limited dynamic range of the PL. To account for this trend, we performed linear fits of both the closed- and open-loop Strehl data, which indicate that the second-stage wavefront control provided a Strehl improvement of $\sim$17\%. Using the extended Maréchal approximation, this corresponds to a reduction in RMS WFE of $\sim$190 nm to $\sim$140 nm, which in turn matches the integral of the difference between the experimentally measured open- and closed-loop PSDs, computed to be 52 nm.

\section{Discussion}\label{disc}

It is encouraging to see that wavelength dispersion already has a beneficial impact on the WFS capability of the PL, even using linear phase retrieval. Since PLs concentrate light on a few pixels and have high optical throughputs, a second-stage PL WFS may also use light more efficiently compared to other cascaded WFS designs. Considering the benefits in precision spectroscopy offered by PLs, we believe that a combined astrophotonic spectrograph and WFS warrants further study. We note some considerations for a such an instrument, as well as WFSing strategies.

In the case of linear phase retrieval, our results suggest that the primary benefit of wavelength dispersion is to improve the spatial resolution of the correction by increasing the number of sensed aberration modes. This result may be somewhat surprising because the space of focal-plane electric fields that couple into the PL depends only weakly on wavelength, at least over our wavelength range. Therefore, we might assume that the space of sensed phase aberrations should have a similarly weak dependence on wavelength. This does not hold because the vector space of focal-plane electric fields is complex and the vector space of phase aberrations is real. For a concrete counterexample, consider a two-mode wavefront sensor that only accepts light in the two orthogonal LP$_{11}$ fiber modes, denoted LP$_{11}$a and LP$_{11}$b. When individually backpropagated to the pupil plane, the phase sampled by each takes the form of a step function that divides the pupil either horizontally or vertically. But the complex linear combination LP$_{11}$a + $i$LP$_{11}$b forms a charge 1 orbital angular momentum mode, which samples a phase vortex. So, the space of sensed phase aberrations depends not only on the shape of the electric field basis vectors (LP$_{11}$a/b), but also the particular linear combination(s) sampled by the wavefront sensor. In the case of the wavelength-dispersed $N$-port PL, different linear combinations are sampled at different wavelengths due to the dispersion of modes as they copropagate through the waveguide. Put another way: at a single, fixed wavelength, a PL senses a small subset of wavefront aberrations (phase \& amplitude) roughly corresponding to a projection of the global wavefront aberration space onto the ``photonic lantern principal modes'' \cite{Kim}. As the wavelength is changed, this modal basis rotates to reveal different projections of the wavefront aberration space \cite{lin:22}, sweeping across the space like a lighthouse beam.

We note two practical limitations to the PLWFS. First, there is a detector tension between spectroscopy, which favors slower readout, and WFSing, which favors faster readout. When where errors evolve slowly (e.g. NCPAs), this tension may be sidestepped, but for a general dispersed WFS this tension motivates high speed detectors with low read noise (e.g. FLI C-RED 1) and brighter targets. Another solution is to divide the outputs of the PL, dispersing some at low spectral resolution for fast WFSing and dedicating others for long-exposure spectroscopy.
 
The second limitation is the dynamic range of the PL WFS, which in our testing could only provide stable correction when the seeing was good ($\sim$ 0.5" or less) and the residual RMS WFE of the primary AO was $\lesssim$ 150--200 nm. While less of an issue for the correction of smaller-amplitude NCPAs, low-wind effect errors can exceed this threshhold, motivating the development of nonlinear phase retrieval methods for wavelength-dispersed, few-moded sensors. Combined with nonlinear methods, wavelength dispersion should lead to an improvement of dynamic range, in the same way that multi-wavelength measurements of a 2$\pi$-wrapped wavefront enable phase unwrapping. These nonlinear methods might use neural networks as in \cite{barnaby} or interpolating methods \cite{Lin:24}. Still, the wavefront sensing benefits offered by spectral dispersion even in the linear regime suggest that dispersed PLs may be well-suited to simultaneously fulfill sensing and spectroscopy functions in future instruments. 

\begin{backmatter}
\bmsection{Funding} Content in the funding section will be generated entirely from details submitted to Prism. Authors may add placeholder text in this section to assess length, but any text added to this section will be replaced during production and will display official funder names along with any grant numbers provided.

\bmsection{Acknowledgment}
The authors would like to thank S. Cunningham, who supported this work as the Subaru telescope operator.  The authors wish to recognize and acknowledge the very significant cultural role and reverence that the summit of Maunakea has always had within the indigenous Hawaiian community, and are most fortunate to have the opportunity to conduct observations from this mountain.
\bmsection{Disclosures}
The authors declare no conflicts of interest.

\bmsection{Data availability}
 Data underlying the results presented in this paper are not publicly available at this time but may be obtained from the authors upon reasonable request.
 
\bmsection{Supplemental document}
See Supplement 1.
\end{backmatter}

\bibliography{sample}

\bibliographyfullrefs{sample}


\ifthenelse{\equal{\journalref}{aop}}{%
\section*{Author Biographies}
\begingroup
\setlength\intextsep{0pt}
\begin{minipage}[t][6.3cm][t]{1.0\textwidth} 
  \begin{wrapfigure}{L}{0.25\textwidth}
    \includegraphics[width=0.25\textwidth]{john_smith.eps}
  \end{wrapfigure}
  \noindent
  {\bfseries John Smith} received his BSc (Mathematics) in 2000 from The University of Maryland. His research interests include lasers and optics.
\end{minipage}
\begin{minipage}{1.0\textwidth}
  \begin{wrapfigure}{L}{0.25\textwidth}
    \includegraphics[width=0.25\textwidth]{alice_smith.eps}
  \end{wrapfigure}
  \noindent
  {\bfseries Alice Smith} also received her BSc (Mathematics) in 2000 from The University of Maryland. Her research interests also include lasers and optics.
\end{minipage}
\endgroup
}{}

\end{document}


\maketitle

\section{Methods}\label{methods}

\subsection{Photonic lantern}
\begin{figure*}
    \centering
    \includegraphics[width=\linewidth]{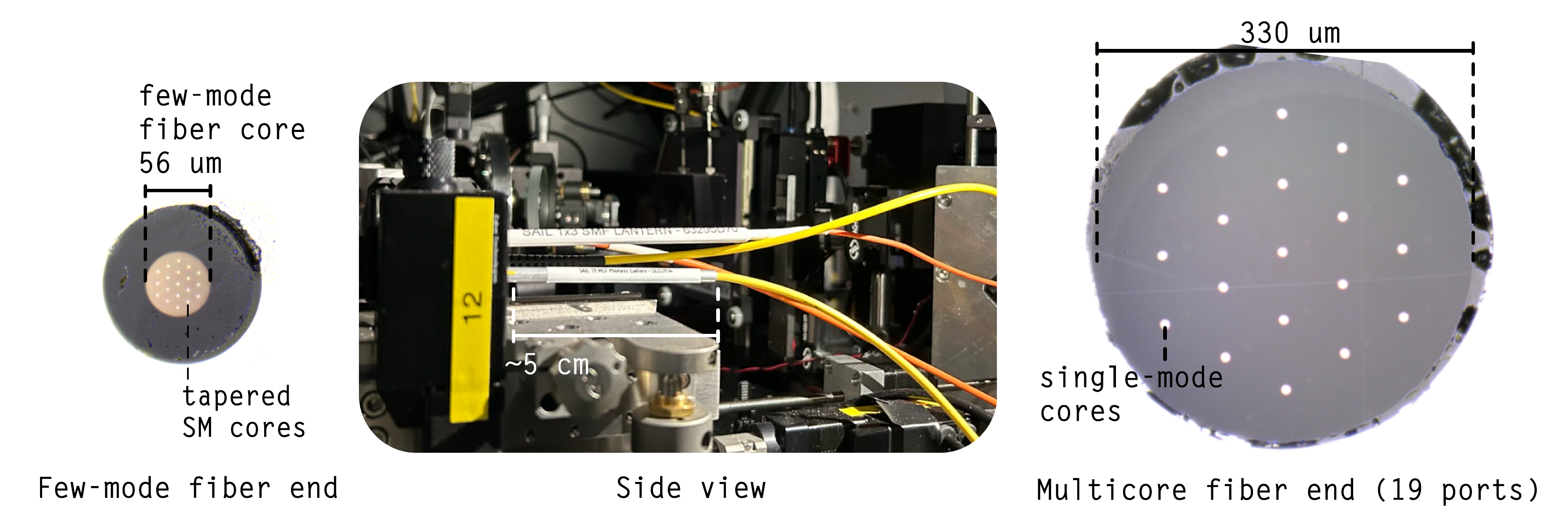}
    \caption{The photonic lanterns currently mounted in the infrared bench of SCExAO. Shown are microscope images of the smaller few-mode fiber and larger output end of the 19-port PL. The middle picture shows the lanterns as mounted within the bench; also shown in the middle is a 3-port PL, which was not used for the experiments in this paper. To make the lantern cores more visible in microscope images of the end faces, visible light was injected into the PL during imaging. The wavelength of the visible light is short enough that it can be confined even in the tapered-down cores at the small end of the PL (left).}
    \label{fig:1}
\end{figure*}

We use a 19-port PL, manufactured by the Sydney Astrophotonics Instrumentation Lab (SAIL), in our wavefront sensing tests. The 19-port lantern was formed from a custom multicore fiber with 19 single-mode channels (hexagonal array, 6.5 $\mu$m core diameter, 60 $\mu$m core spacing, numerical aperture 0.14); this fiber was inserted into a lower-index fluorine glass capillary, and then one end was heated and drawn to form a tapered structure with an entrance diameter of 56 $\mu$m. According to these parameters, the outputs of this lantern are single-moded for $\lambda>$ 1200 nm.  
Figure \ref{fig:1} gives microscope images for both endfaces of the lantern, as well as a picture of the lantern as mounted within SCExAO. To make the lantern cores more visible, they were illuminated with visible light during microscope imaging. This PL has a maximum measured coupling efficiency of 80\%, which was obtained at $\lambda = 1550$ nm.

\subsection{SCExAO testbed} \label{sec:setup}
All experiments in this paper were performed on the infrared bench of the SCExAO testbed\cite{Jovanovic_SCE2015}. The top panel of Figure \ref{fig:2} gives a simplified overview of the infrared bench, showing only optical components immediately relevant to our experiments. Light is emitted from a supercontinuum laser source (NKT superK) which is collimated by an off-axis parabolic mirror. The collimated beam reflects off SCExAO's $50\times 50$ actuator MEMS deformable mirror, and then is optionally apodized to a near-Gaussian beam profile using a pair of beam-shaping lenses. These lenses were used for daytime tests because some of the downstream optics were designed for the smaller, shaped beam; for later on-sky testing, we tweaked the alignment of the setup to omit these lenses and simplify the beam path. The lenses do not appear to have a strong impact on the WFSing ability of our system. From here, the beam is intercepted by a 90:10 splitter which directs the majority of the light to the injection platform, a four-axis movement stage holding a lens and the 19-port PL. The movement stage controls the focal ratio of the beam and the transverse alignment relative to the lantern (for more details about the injection see~\citenum{Jovanovic_EIL2017}). The remaining 10\% of light is directed to an internal InGaAs camera (FLI C-RED 2) for PSF monitoring. Finally, the output of the PL is routed to a diffraction-limited spectrograph with $R\approx$ 700, 500, and 300 in the astronomical $y$, $J$, and $H$ bands, respectively; the spectral traces of the lanterns are imaged either onto an FLI C-RED 1 (320 $\times$ 256 pixels, 24 $\mu$m HgCdTe e-APD) or C-RED 2 detector (640x512 pixels, 10 $\mu$m pixel InGaAs) --- our setup occasionally changed between experiments due to external constraints. All plots in the main manuscript show data taken when our setup used the C-RED 1 detector.
\\\\
Because the output of the 19-port PL is a multicore fiber, it is dispersed in a TIGER configuration \cite{sergio:22}; the narrow core-to-core spacing of this particular lantern makes the 19 spectral traces partially overlap. As far as we can tell, this does not impact wavefront sensing, though it would impact spectroscopic applications. Another complication is that the shortest wavelength sampled by the spectrograph is $\sim$1100 nm, shorter than the single-mode cutoff wavelength for the PL outputs. Due to the TIGER dispersion and the overlap of the spectral traces, we were unable to remove wavelength information below this cutoff without sacrificing light in the correct wavelength range from other ports, implying that the PL was not acting exactly as a multi-mode to single-mode converter. That said, barring manufacturing defects, we would still expect the PL outputs at these wavelengths to be close to the LP$_{\rm 01}$ mode, which matches our observations. Due to the tapered structure of the PL, any wavefront propagating through the PL will initially couple into cores that are small enough to be single-moded. Additionally, since propagation through a PL is often assumed to be adiabatic, any light initially in the fundamental mode of a core will remain in the fundamental mode. To make this argument more rigorous, we use the \texttt{cbeam} \cite{cbeam} package to compute the coupling coefficients of a tapered optical fiber whose parameters match a single tapered core of our lantern, in the region where the cores become multi-moded. Under the simplifying assumption that our lantern has a linear taper profile, we find that the coupling coefficient between the LP$_{\rm 01}$ mode and the LP$_{\rm 11}$ mode group is of order $10^{-4}$ cm$^{-1}$. A comparison of the corresponding length scale with that of our PL implies that cross-coupling is negligible.
\subsection{Strehl calculation}
Strehl is estimated from PSFs recorded by SCExAO's PSF monitoring camera, which is operated simultaneously with the PL wavefront control loop. We perform a basic Strehl extraction. Given a time series of aberrated PSFs, we first compute the average of the series and locate the point of maximum flux. We then normalize each PSF and compare the flux at this point to the maximum flux of a normalized and ``unaberrated'' PSF, measured experimentally during the day using SCExAO's calibration source. Of course, due to small optomechanical drifts our calibration PSF is not completely unaberrated, slightly biasing our extracted Strehl ratios upwards. Given the performance of SCExAO, we expect this bias to be $1-2\%$ at most, with negligible impact on relative comparisons between Strehl ratios. Such comparisons are more relevant to gauge the on-sky performance of the PLWFS.

\subsection{Software}
\begin{figure}
    \centering
    \includegraphics[width=\textwidth]{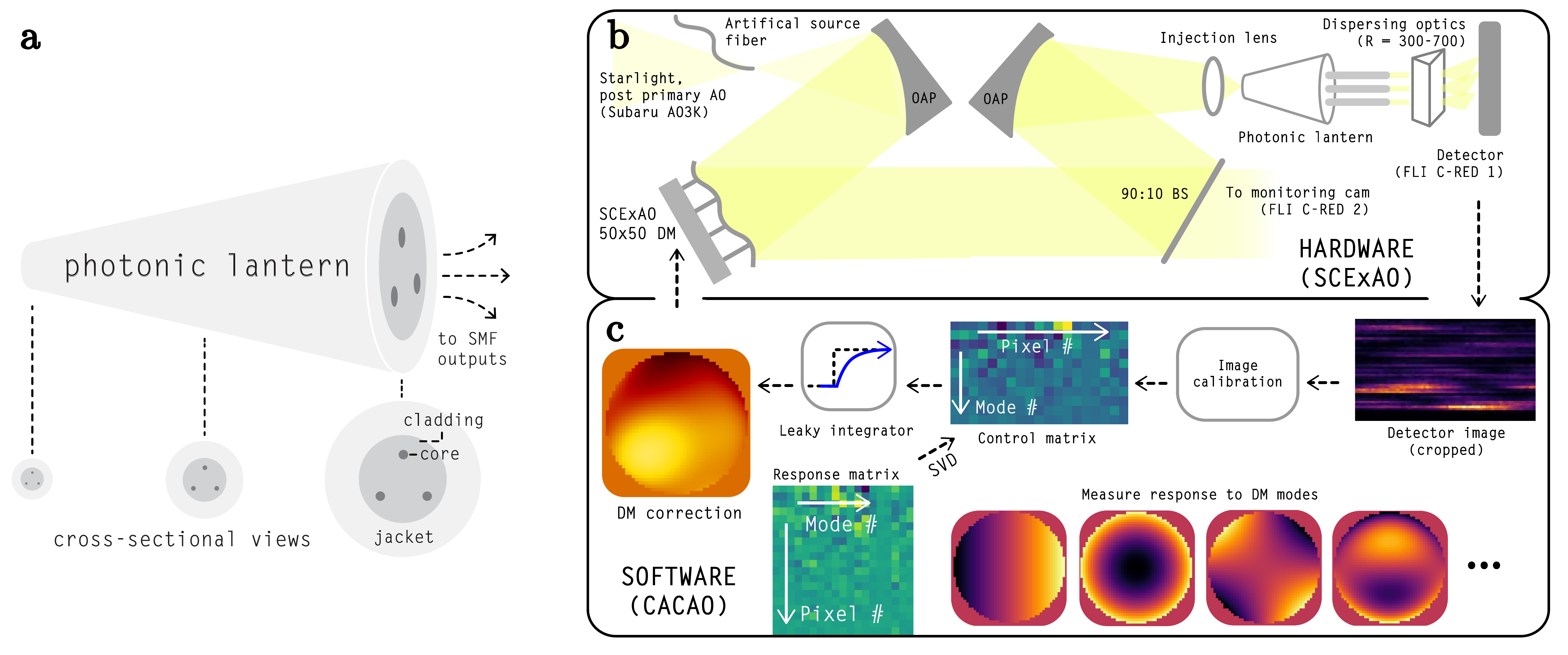}
    \caption{The same as Figure 1, repeated for reference. \textbf{a}: A 3-port PL, with 3 single-moded outputs. The PL spatially multiplexes light propagating in multiple fiber modes into separate single-moded outputs. \textbf{b}: simplified beam diagram of the astrophotonics platform at SCExAO/Subaru. Light, either from an astronomical source or an supercontinuum white light laser, is collimated by an off-axis parabola (OAP). The beam passes through a 2500-actuator deformable mirror, and may optionally be apodized into a Gaussian beam profile (instead of the flat-topped beam profile produced by a point source) using a pair of beam-shaping lenses. A beamsplitter redirect 90\% of the light to the astrophotonics injection unit, which contains an injection lens and 19-port PL mounted on a 4-axis stage controlling transverse alignment of the PL relative to the beam and the focal ratio. The outputs lantern are dispersed at low spectral resolution and imaged on an FLI C-RED 1 or C-RED 2, depending on external constraints. \textbf{c}: an overview of the software steps involved in closing the wavefront control loop. All steps were performed using the CACAO package.}\label{fig:2}
\end{figure}

We use the Compute And Control for Adaptive Optics (CACAO; \cite{CACAO}) package to calibrate the wavelength-dispersed PLWFS and run the wavefront correction loop. An overview of the software is given in the bottom panel of Figure \ref{fig:2}. We process the WFS image in a number of ways before passing it to the control loop, involving cropping the image around the lit area of the detector and the usual steps of dark subtraction, masking bad pixels, image normalization, and WFS reference image subtraction. For closed-loop operation, we use a linear reconstructor and a leaky integrator controller, detailed in the next subsection.

\subsection{WFS calibration}
We calibrate and operate dispersed PLWFS in the standard way for astronomical AO, which we review below. First, we measure the sensor response to an incomplete modal basis of pupil-plane phase aberrations --- our ``calibration basis''. Common choices include the Zernike, Fourier, and zonal (DM actuator) modes. We column-stack the measured response to each of the $N$ calibration modes into an $M \times N$ response matrix, denoted $A$, where $M$ is the total number of pixels in the WFS intensity image. Next, we apply a singular value decomposition to $A$. When computing the control matrix (pseudoinverse of $A$) we choose to only keep singular values that were at least on-tenth the largest value. Finally, we drive the SCExAO DM using a leaky integrator controller: the mirror control signal $\bm{\phi}$ at step $n+1$ is set as
\begin{equation}
        \bm{\phi}_{n+1} = l \bm{\phi}_{n} - g A^+ \left(\bm{I}_n -\bm{I}_0\right).
\end{equation}
Here, $\bm{I}$ is the intensity image of the WFS, $\bm{I}_0$ is the reference intensity, $g$ is the gain, and $l$ is the leak. The same loop parameters are used for all modes.

\section{Off-sky testing with zonal basis calibration}\label{sec:zonal}
\begin{figure}
    \centering
    \includegraphics[width=0.8\linewidth]{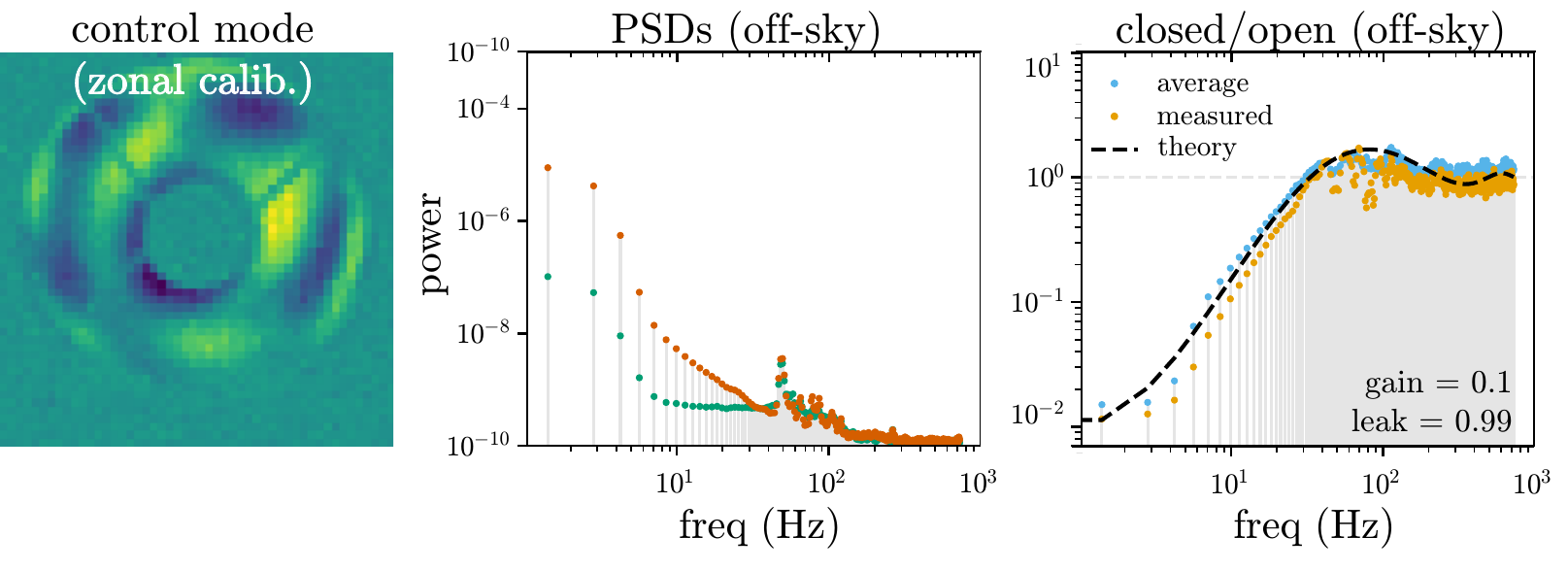}
    \caption{Wavefront control performance of the dispersed PL WFS during daytime laboratory tests. Left: a control mode in our wavefront control loop; the sensor was calibrated in zonal basis. Middle: Open- and closed-loop PSDs of this mode's amplitude. Right: the ratio of the closed and open loop PSDs, which estimates the squared modulus of the rejection transfer function. Orange dots correspond to the given mode, and blue to the average over all modes; the dashed black lines give the expected transfer function for our control parameters.}
    \label{fig:3}
\end{figure}

In addition to the off-sky wavefront control test from the main text, where the PL WFS was calibrated against the first 100 non-piston Zernike modes, we also tried a calibrating in the zonal mode basis, which is more typical for astronomical AO. Zonal modes correspond to the pistoning of each DM element. This calibration yielded 58 control modes; we then closed the AO loop using a leaky integrator controller with a leak of 0.99 and a gain of 0.1. Due to external constraints, the PL was dispersed onto an FLI C-RED 2 camera instead of the C-RED 1 used in previous tests; we ran this camera at a framerate of 1450 Hz and measured a system latency of 2.96 frames. Similar to the earlier demonstration which used Zernike basis, we then tested the system by injecting high-pass-filtered Kolmogorov phase screens with an RMS amplitude of 100 nm and a windspeed of 10 m/s onto a separate channel of the DM. From this test, we obtained the open- and closed-loop PSDs and the rejection transfer function, as shown in Figure \ref{fig:3}. We note that the agreement between the experimentally measured average transfer function and the theoretical expectation is somewhat better in this test than the prior test, but it is unclear if this discrepancy is due to the difference in calibration basis, the detector swap, optomechanical instability of the SCExAO system, or some mixture of the three.

\section{Reference spectrum compensation}\label{secA1}

If the calibration of the dispersed WFS is dependent on the reference spectrum used for calibration, then such a sensor would require recalibration every time the calibrator source spectrum changes. We consider how changes in source spectrum may be accounted for in the linear model, in the case of aberrations which are achromatic in terms of optical path difference (OPD). First, note that the power $\bm{p}$ measured through the WFS at some phase aberration $\bm{r}$ will be proportional to the source spectrum $s(\lambda)$:
\begin{equation}
    p_j(\bm{r},\lambda) = f_j(\bm{r},\lambda) s(\lambda).
\end{equation}
Here, $f_j$ is a scalar-valued function representing the wavelength-dependent throughput of sensor output $j$, which is intrinsic to the WFS. Next, discretize over wavelength, yielding 
\begin{equation}
\begin{split}
    p_{ij} - p_{ij,0} &= \sum_k A_{ijk} x_k \\
    p_{ij,0} &= f_j(\bm{r},\lambda_i) s(\lambda_i) \\
    p_{ij} &= f_j(\bm{r}+\bm{x},\lambda_i)  s(\lambda_i) 
\end{split}
\end{equation}
where $i$ iterates over wavelengths and $j$ iterates over output ports; the matrices $A_{1jk},A_{2jk}...$ are the linear response matrices in each spectral channel. The reference intensity measured using the calibration source is $\bm{p}_{ij,0}$, and $\bm{x}_k$ now represents OPD. The sensor ``slopes'' in $A_{ijk}$ are linearly proportional to the absolute power in spectral channel $i$. Thus, during the calibration process we should factor out the source spectrum, defining the normalized response matrix $A'$ as
\begin{equation}
    \dfrac{p_{ij}}{p_{ij,0}} - 1 = \sum_k \dfrac{1}{p_{ij,0}} A_{ijk} x_k \equiv \sum_k  A_{ijk}' x_k.
\end{equation}
Wavelengths $\lambda_i$ for which $p_{ij,0}$ are small should be discarded, since they will amplify noise; as a corollary, it may be preferable to choose a calibration source with the flattest possible spectrum over the sensor's wavelength band. During operation, the sensor may see a different reference spectral response, $p_{ij,\star}$. In this case, the control law is obtained by inverting 
\begin{equation}\label{eq:guide}
    p_{ij} - p_{ij,\star} = \sum_k p_{ij,\star} A_{ijk}' x_k.
\end{equation}
WFS image normalization can be accounted for by a rescaling of $p_{ij,\star}$. The above implies that the control matrix should be recomputed whenever the source spectrum changes, corresponding to the new spectral reference. However, the spectro-normalized response matrix $A'$ does not need to be recomputed unless the instrument is unstable. We have confirmed that this method works by applying different passband filters and closing the loop using the 3-port PL on SCExAO, with the same normalized response matrix. In practice, it might be desirable to live-update both the reference response $p_{ij,\star}$ and the control matrix. Alternatively, one may choose to use the spectrally normalized signal $p_{ij}/p_{ij,\star}$ in conjunction with $(A')^+$ as the control matrix; however, since spectral normalization will re-scale noise, the predicted phase will no longer be the maximum likelihood estimate. This is especially problematic if there regions where $p_{ij,\star}$ is near 0 (e.g. atmospheric absorption features). We leave a more rigorous performance comparison between different spectral compensation methods for future work.
\\\\
If the calibration spectrum is somehow similar to the target guide star spectrum, then we might approximately solve equation \ref{eq:guide} as
\begin{equation}
    x_k \approx  \sum_{ij} \dfrac{p_{ij,0}}{p_{ij,\star}}A^+_{ijk}\left( p_{ij} - p_{ij,\star}\right)
\end{equation}
which is reminiscent of how the sensitivity loss of the pyramid WFS in the presence of small phase offsets is compensated (``optical gain compensation'', e.g. \citenum{chamb}). Therefore, this effect might be called ``spectral gain compensation'', though as a name, ``reference spectrum compensation'' is probably more informative. Whereas optical gain arises from a phase offset combined with WFS nonlinearity \cite{vince}, ``spectral gain'' arises from a spectral offset coupled with the {\it linear} dependence of the WFS slopes on the source spectrum.

\bibliography{sample}
